# Transmission of low energy electrons through a polyethylene terephthalate 800-nm diameter nanocapillary [*]

LI Pengfei[1, 2], LIU Wanqi[1], HA Shuai[2], PAN Yuzhou[2], FAN Xuhong[2], DU Zhanhui[2], WAN Chengliang[2], CUI Ying[2, *], YAO Ke[3, 4], MA Yue[5], YANG Zhihu[6], SHAO Caojie[6], Reinhold Schuch[7], LU Di[8], SONG Yushou[1], ZHANG Hongqiang[2, *], CHEN Ximeng[2]

1.College of Nuclear Science and Technology, Harbin Engineering University, Harbin 150000, China

2.School of Nuclear Science and Technology, Lanzhou University, Lanzhou 730000, China

3.Institute of Modern Physics, Fudan University, Shanghai 200433, China

4.Key Laboratory of Nuclear Physics and Ion-Beam Application (MOE), Fudan University, Shanghai 200433, China

5.RIKEN Nishina Center, RIKEN, Wako 351-0198, Japan

6.Institute of Modern Physics, Chinese Academy of Sciences, Lanzhou 730000, China

7.Physics Department, Stockholm University, Stockholm S-10691, Sweden

8.Department of Physics, University of Gothenburg, Gothenburg S-40530, Sweden

**Abstract**

The transmission of 2-keV electrons through a polyethylene terephthalate (PET) nanocapillary with a diameter of 800 nm and a length of 10 μm is studied. The transmitted electrons are detected using a microchannel plate (MCP) with a phosphor screen. It is found that the transmission rate for the transmitted electrons at the incident energy can reach up to 10% for an aligned capillary in the beam direction, but drops to less than 1% when the tilt angle exceeds the geometrically allowable angle. The transmitted electrons with the incident energy do not move with changes in the tilt angle, so the incident electrons are not guided in the insulating capillary, which is different from the behavior of positive ions. In the final stage of transmission, the angular distribution of the transmitted electrons within the geometrically allowable angle splits into two peaks along the observation angle perpendicular to the tilt angle. The time evolution of the transmitted complete angular distribution shows that when the beam is turned on, the transmission profile forms a single peak. As the incident charge and time accumulate, the transmission profile begins to stretch in the plane perpendicular to the tilt angle and gradually splits into two peaks. When the tilt angle of the nanocapillary exceeds the geometrically allowable angle, this splitting tends to disappear. Simulation of charge deposition in the capillary directly exposed to the beam indicates the formation of positive charge patches, which are not conducive to guiding, unlike positive ions. Based on the simulation results, we can explain our data and discuss the possible reasons for the splitting of the transmission angular profiles.

# 1. Introduction

The discovery of the guiding effect began in 2002 with the study of the transport of charged particles in insulating nanocapillaries [1]. When slow, highly charged ions (3 keV-$Ne^{7+}$) are transported through an insulating membrane (polyethylene terephthalate, PET) with nanocapillaries, even if the angle between the normal direction of the membrane plane and the direction of the incident ion beam (referred to as the tilt angle) reaches 20°, most of the emitted ions retain their initial energy and charge state. This provides a possibility for the development of new charged particle beam optics. Early studies on the transport of positive ions in insulating nanocapillaries and single glass capillaries [1-19] revealed the mechanism of the guiding effect: the interaction between positive ions and the inner walls of capillaries causes positive charges to deposit on the inner walls, forming charge spots through self-organization, which generates a Coulomb field that repels subsequent incident ions, allowing subsequent positive ions to pass along the axis of capillaries without impacting the inner wall, thus avoiding charge exchange and energy loss during the transport process [10,11].

At present, the physical mechanism of positive ion transportation has been studied clearly, and the effects of Coulomb field of deposited charge spot, image charge force and near-surface scattering on the transport of[18–20] have also been revealed. The effect of image charge force[13,14] on ion beam shaping has been found in the experiment of $Ne^{7+}$ ions passing through mica microporous membranes with rhombic and rectangular holes. In the experiment of 70 keV $Ne^{7+}$ ions passing through a mica microporous membrane with diamond-shaped pores, a banana-like image of transmitted neutral particles was observed, indicating that the scattering process plays an important role in ion transmission for higher energy ions [15].

At present, the physical mechanism of positive ion transportation has been clearly studied [18-20]. The roles of the Coulomb field of deposited charge spots, image charge force, and near surface scattering in the transport process have also been revealed. In the experiment of $Ne^{7+}$ ions transmitting through mica membranes with capillaries of rhombic and rectangular cross-sections, it was found that the image charge force [13,14] can shape the ion beam. In the experiment of 70 keV $Ne^{7+}$ ions transmitting through a mica membrane with capillaries of rhombic cross-section, a banana shape formed by the transmission neutral particles was observed, indicating that, for high-energy ions, the scattering process plays an important role in transport [15].

However, the transport mechanism of negative charged particles in insulating capillaries is still controversial [21–45]. When negative ions with energy above 10 keV transmit through insulating capillaries [21–26], both the scattering process and the charge exchange process play major roles. The angular distribution of transmitted ions shows a bimodal structure: the negative ions form a peak along the initial beam direction, while the scattered neutral particles and positive ions form another peak along the axial direction of the capillaries. At present, the negative ions studies are mainly those with energies greater than 10 keV, and charge deposition does not play a major role in the particles transport. There is no experimental data for negative ions with energies below 10 keV, and whether they have a guiding effect similar to that of positive ions has not yet been concluded. At the same time, researchers have conducted numerous studies on the transmission of low-energy electrons in the energy range of 0.05–2 keV through insulating capillaries [27–45]. The experiments on low-energy electrons transmission through $Al_2O_3$ membrane with nanocapillaries show that[27–29] elastic scattering electrons are emitted along the axial direction of the nanocapillaries, and the electron transmission rate decreases with the accumulation of incident charge (charging time). In the experiments of low-energy electrons transmission through PET membrane with nanocapillaries [30–35], only the electrons near the initial energy were measured. At the beginning of charging process, there were almost no transmitted electrons. After a period of silence (several minutes to tens of minutes), the intensity of transmission electrons increased rapidly, reached its highest value, and then decreased slowly, which could be regarded as stable. The energy of the transmitted electrons oscillates rapidly in the rapid rise stage of the intensity and remains at a value slightly smaller than the initial energy[32,33] in the stable stage of the intensity. The simulation of electron transport [31] shows that the deposited charges do not play a major role in this process, and the main transmission characteristics are caused by the electron scattering. The calculation of the transmission rate shows that the transmission rate decreases with the increase of the incident charge, which is consistent with the experimental results of low-energy electron transmission through $Al_2O_3$ nanocapillaries [29], but inconsistent with the experimental results for PET nanocapillaries [32,33]. In terms of conductivity, there is only a slight difference of the volume conductivity between PET and $Al_2O_3$; In terms of membrane structure, there is little difference in the parameters of nanocapillaries between PET membrane and $Al_2O_3$ membrane used in the literature. The factors that cause the large difference in the charging process between the two need to be studied.

In the above study of negative charged particles [25-30,32-40], a one-dimensional electrostatic energy analyzer was used in the experiment. Due to the limitation of the entrance slit, it can only measure the one-dimensional angular distribution by changing the detection angle of the detector and cannot measure the evolution of the full angular distribution of transmitted particles over time during the charging process. In order to study the controversial issue of how charge deposition affects the transport of low-energy electrons in insulating nanocapillaries, a two-dimensional position-sensitive detector composed of the microchannel plates and a phosphor screen was used to measure the two-dimensional angular distribution of 2 keV electrons transmission through PET nanocapillaries and its evolution with charging. It was found that the angular distribution of transmitted electrons splits into two electron spots in the observation direction perpendicular to the tilt angle, and the two electron spots approach each other and eventually coincide as the tilt angle increases. The simulation of charge deposition caused by electrons shows that low-energy electrons cause the emission of a large number of secondary electrons, which leads to a high-density hole region with positive charges within 1 nm of the material surface. Meanwhile, the incident electrons have a greater probability of depositing within tens of nanometers below the material surface, forming a negative charge region. The deposition of charges with opposite polarities leads to a complex charging behavior of transmitted electrons. The possible reason for the splitting of the angular distribution of the transmitted electrons is also discussed.

## 2. Experimental equipment

The experiments were performed on the low-energy charged particle beam experimental platform of Lanzhou University. A schematic diagram of the experimental setup is shown in Fig. 1. The electron beam is generated by a $LaB_6$ electron gun. To prevent the detector from collecting the photons generated by the filament, a 90 ° electrostatic deflector is used to deflect the electron beam. The beam direction is then adjusted in the horizontal and vertical directions by the electrostatic deflector, focused by the Einzel electrostatic lens, and finally collimated by a pair of adjustable slits and a solenoid coil into the target chamber. The target chamber is made of μ metal, which effectively shields the external static magnetic field. The PET membrane target is mounted on a five-dimensional adjustment device, which allows movement in three spatial dimensions and rotation in horizontal and vertical planes. After transmitting through the membrane, the electrons pass through an electron energy analyzer composed of three grids and are detected by a two-dimensional position-sensitive detector made of MCPs and a phosphor screen. 3D Homhertz coils are placed around the experimental setup to counteract the geomagnetic field. The vacuum in target chamber is maintained at better than $3 \times 10^{-8}$ mbar (1 bar = $10^5$ Pa) during the experiments.

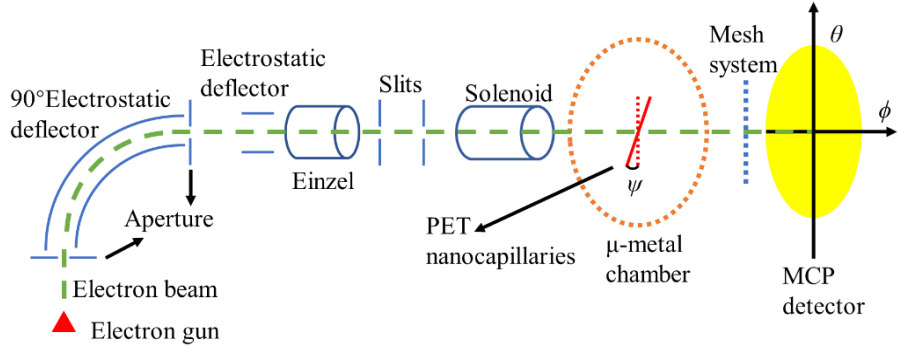

**Figure 1.** A schematic drawing of the experimental setup, the tilt angle $\psi$ between the axes of capillaries and the electron beam, the observation angles $\phi$ and $\theta$ given with respect to the electron beam are defined.

In the experiment, the angle corresponding to the maximum electron transmission intensity was determined by rotating the PET membrane in two rotational dimensions, with the axis of nanocapillaries aligned with the electron beam's incident direction. Therefore, this angle was defined as 0 °. The tilt angle $\psi$ is the angle between the axis of the nanocapillaries and the primary electron beam. The observation angles $\phi$ and $\theta$ are defined with respect to the electron beam incident direction, where $\phi$ and $\psi$ lie in the same rotational plane, and $\theta$ lies in the plane perpendicular to $\psi$.

The PET membrane was irradiated with high-energy ions to produce tracks and then etched with NaOH solution to transform the tracks into nanocapillaries, thus obtaining the PET nanocapillaries used in the experiments. The membrane thickness is 10 μm, the pore diameter is 800 nm, and the corresponding geometric opening angle, calculated from the aspect ratio, is 4.6 °. The density of nanocapillaries in the membrane is $2.4 \times 10^7$ mm$^{-2}$, the average spacing between nanocapillaries is 1.2 μm, and the axial divergence angle of the nanocapillaries is less than 0.5 °, corresponding to a geometric transmission rate of 12.3%. The volume conductivity of PET is $1.0 \times 10^{-16}$ Ω$^{-1}$ · m$^{-1}$ [46], the surface conductivity is $1.0 \times 10^{-18}$ Ω$^{-1}$ · m$^{-2}$ [46], and the dielectric constant is about 3.3[46]. The front and back surfaces of the membrane are coated with 10 nm thick gold layers to prevent surface charging.

The MCP detector was used to measure the primary beam of the 2 keV electron beam. By keeping the parameters of the beam transport unchanged and reducing the filament heating power, the electron beam intensity was maintained within the allowable range of the MCP detector. The two-dimensional angular distribution image of the electron primary beam and its projection on the two planes of $\phi$ and $\theta$ are shown in Fig. 2, and the beam spot size is 1.0 mm × 1.0 mm. The electron beam is measured using a Faraday cup on the target holder. Increasing the filament heating power increases the electron beam intensity to the order of pA. When the beam current reaches a steady state, the beam spot size measured by the Faraday cup is 2.0 mm × 1.8 mm, and the intensity is -0.95 pA/mm$^2$, meaning 3 electrons enter a single nanocapillaries per second (3$e$/capillary/s). The divergence angle of the electron beam, calculated from the beam diameters at the target and detector positions, is less than 0.3 °. The geometric transmission angle of the electron beam through the membrane is defined as in our previous work[24], that is, $\sigma_{geot} = \sqrt{\sigma_{asp}^2 + \sigma_{beam}^2 + \sigma_{axis}^2}$, where $\sigma_{asp}$, $\sigma_{beam}$ and $\sigma_{axis}$ are the geometric opening angle (4.6 °), the electron beam divergence (0.3 °), and the axial divergence (0.5 °) of the nanocapillaries, respectively. The calculated geometric transmission angle is about 4.6 °, which is approximately equal to the geometric opening angle of the nanocapillaries.

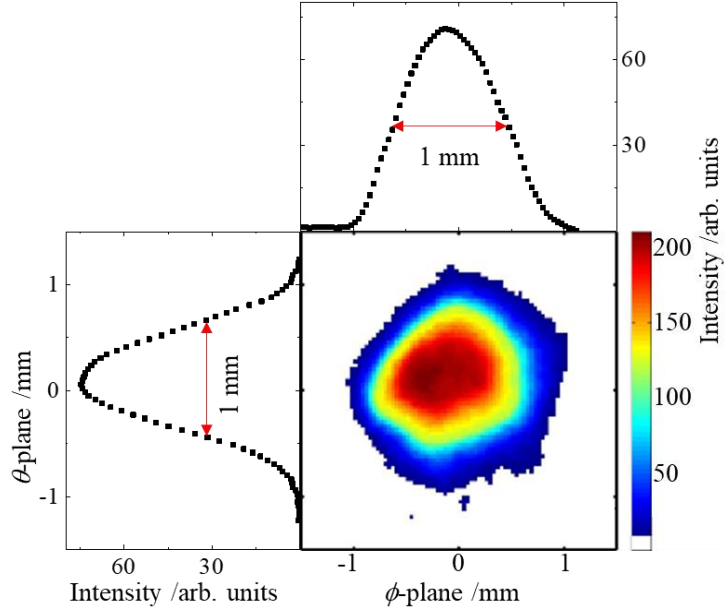

**Figure 2.** A primary beam profile of 2 keV electrons and the corresponding projections on the planes of $\phi$ and $\theta$.

## 3. Experimental result

By measuring the energy distribution of transmitted electrons, it was found that the energy distribution of transmitted electrons is essentially the same as that of incident electrons. To avoid the influence of secondary electrons (the sources of secondary electrons are more complex [24]), we focused on the transmitted electrons with incident energy. In the subsequent measurements, a voltage of -20 V was applied to the central grid of the electron energy analyzer to filter out the low-energy secondary electrons.

The two-dimensional angular distribution images of 2 keV electrons transmitted through PET nanocapillaries at different tilt angles ($\psi$) when the charge reaches a steady state are shown in Fig. 3. Each image is accumulated by 300 $e$/capillary of incident charge, corresponding to a time duration of 100 s. At the observation angle ($\phi$), the center of the transmitted electron angular distribution shifts slightly with the tilt angle. Within the geometric opening angle, the image splits into two spots along the observation angle ($\theta$). As the tilt angle ($\psi$) increases from 0 °, the two electron spots become smaller in size and intensity, move closer together, and finally merge into a single electron spot when the tilt angle ($\psi$) is approximately equal to the geometric transmission angle ($\pm$ 5 °). As the tilt angle ($\psi$) continues to increase, one electron spot remains on the transmission angle distribution image and continues to shrink, eventually disappearing. When the tilt angle moves to the positive side, the center of the angular distribution of the transmitted electrons shifts to the negative observation angle; when the tilt angle moves to the negative side, the center of the angular distribution of the transmitted electrons shifts to the positive observation angle.

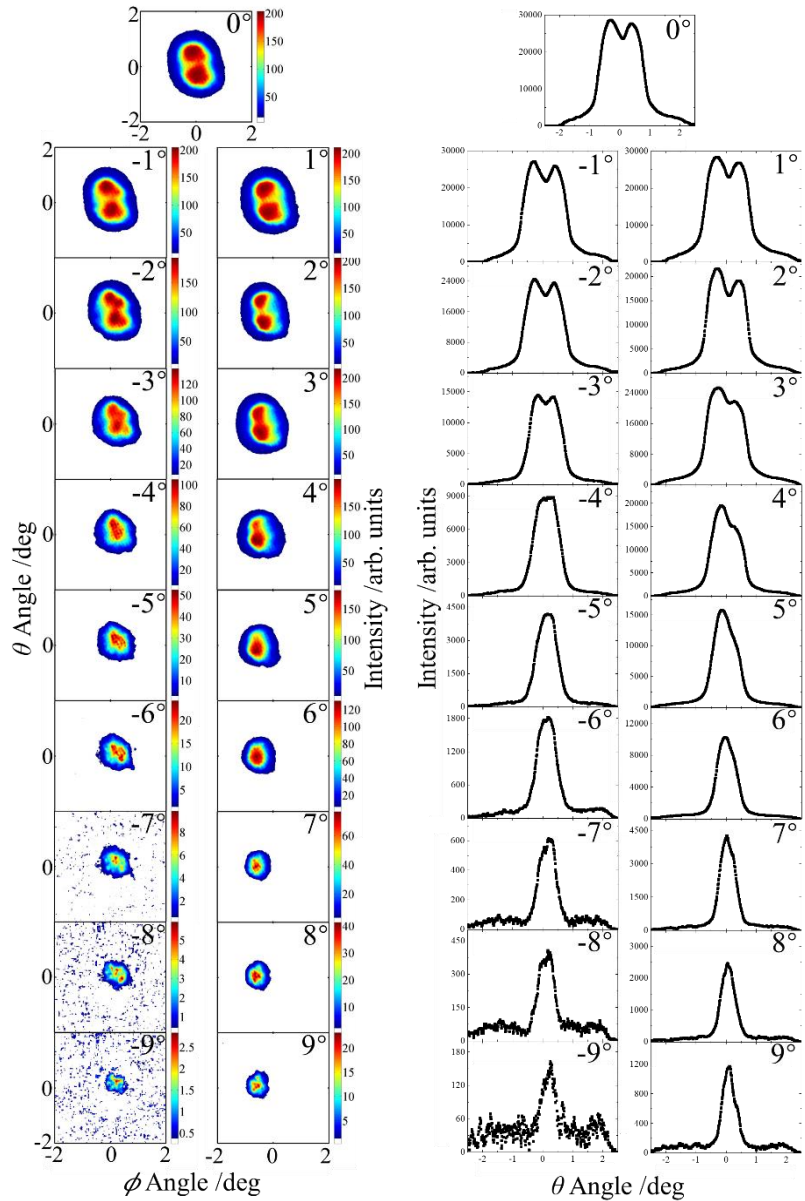

**Figure 3.** Two-dimensional transmission angle distribution of 2 keV electrons through PET nanocapillaries in a stationary state at different tilt angles: (a) Two-dimensional transmission angle distribution; (b) $\theta$ plane transmission angle distribution projection.

The transmission rate of 2 keV electrons as a function of the tilt angle ($\psi$) at steady state is shown in Fig. 4. The angle between the red dashed lines represents the geometric transmission angle. When the tilt angle is 0 °, the electron transmission rate can reach 10%. When the tilt angle is smaller than the geometric transmission angle, the electron transmission rate decreases rapidly with the increase in the tilt angle. When the tilt angle exceeds the geometric transmission angle, the transmission rate increases slowly with the tilt angle. The charging process of electrons transmitted through the PET membrane was measured at a tilt angle of 0 °. The time evolution of the full angular distribution of transmitted electrons is shown in the Fig. 5(b). Prior to the measurement, the PET membrane was fully discharged. Each image in the figure corresponds to an accumulated incident charge of 80$e$/capillary and a time duration of 27 s. The image shows that there is only one transmitted electron spot at the beginning of the charging process. As the incident charge accumulates (charging time increases), the electron spot begins to stretch along the $\theta$ direction and gradually splits into two transmitted electron spots. The electron transmission rate as a function of charging time is shown in Fig. 5(a). During the charging process, the transmission rate oscillates around 9%.

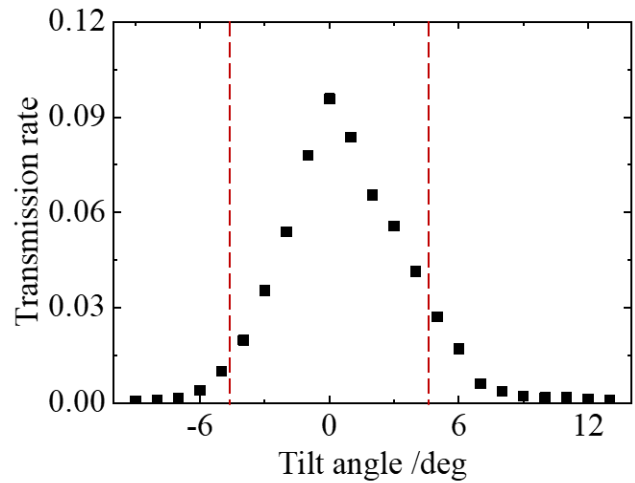

**Figure 4.** The transmission rate of 2 keV electrons in stationary state as a function of $\psi$, the red dash lines stand for the geometrical transmission angle.

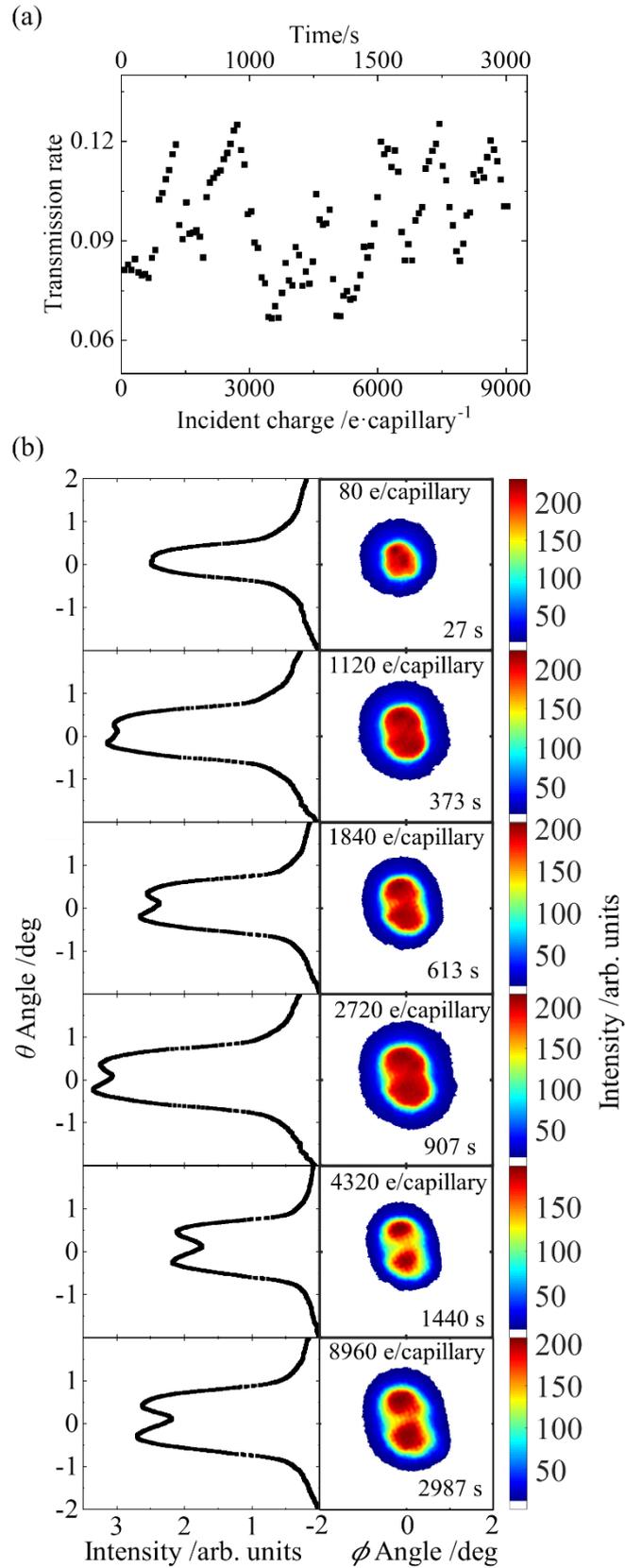

**Figure 5.** The time evolution of transmitted angular distributions of 2 keV electrons at the tilt angle of 0° through PET nanocapillaries during the charging process: (a) The evolution of electron transmission rates; (b) projections of the transmitted angular distributions.

Gaussian fitting of the projection of the transmission angle distribution on the $\phi$ and $\theta$ planes was performed. In the $\phi$ plane, the peak position shifts, and the full width at half maximum (FWHM) oscillates synchronously with the transmission rate. The peak position starts at -0.12 ° and approaches 0 ° after oscillation, while the FWHM oscillates and broadens. On the $\theta$ plane, the evolution of the upper and lower peak positions is shown in Fig. 6(a) and (b). As the incident charge accumulates, the upper spot peak position shifts to about 0.5 °, while the lower spot peak position shifts to about -0.3 °. The corresponding FWHMs of the two peaks oscillate around 0.6 °. Similarly, the peak position shifts and the FWHM oscillates synchronously with the transmission rate.

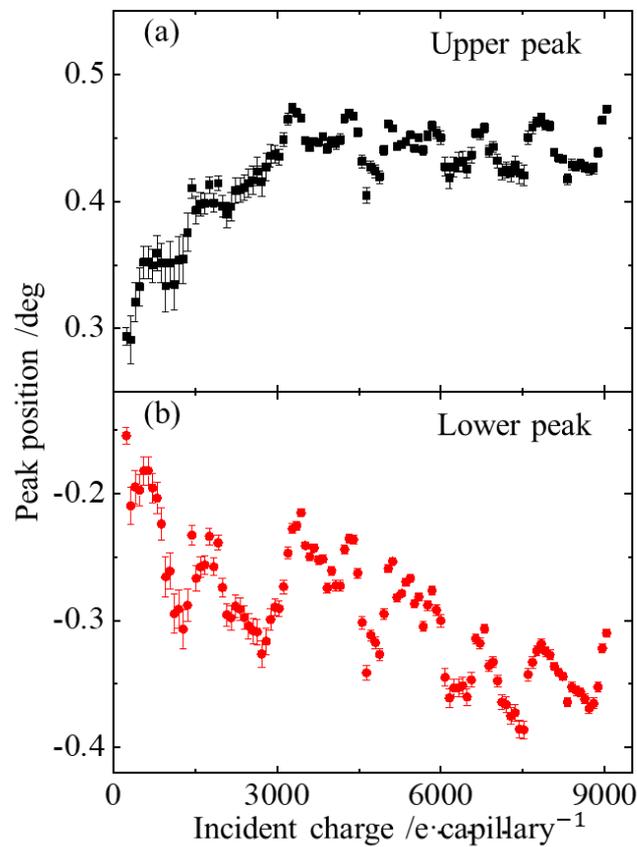

**Figure 6.** At tilt angle 0°, as the charging time accumulates, the evolution of upper (a) and lower (b) peak positions on plane, respectively.

## 4. Discussion

Unlike the guiding effect of positive ions, electron transport in insulating nanocapillaries involves more complex physical processes. For example, when the tilt angle exceeds the geometric opening angle, electrons scattering become the dominant process. However, there is still much controversy surrounding the process of charge deposition on the inner wall of nanocapillary caused by incident electrons [27–45]. For nanocapillaries made from different insulating materials, the measurement results of the electron transmission rate as a function of time indicate different polarities of the deposited charge. Additionally, there are discrepancies between the theoretical simulation and the experimental results of electron transport [27–34]. The collision of electrons with the inner wall of the nanocapillary generates a large number of secondary electrons, further complicating the charging process. To study the charge deposition on the inner wall of nanocapillary caused by electrons, the physical processes of electron scattering, secondary electron excitation, and incident electron deposition were considered. The charge deposition and its spatial distribution on the inner wall of nanocapillary caused by 2 keV electrons were calculated using "CASINO" software [47–49].

In CASINO, the Monte Carlo method is used to calculate the electron track in detail, and the physical process between the electron and the sample are modeled using random numbers and probability distributions [48]. The electro tracking ends when the electron is either emitted from the sample or captured by the sample (i.e., when the electron's energy drops below a certain threshold). In the simulation of secondary electrons, the work function of the sample is used as the threshold. The energy loss during each step of electron transport includes the excitation of plasmons, inter- and intra-band transitions, and core excitations. When the incident electron energy is low, the electron energy loss in the sample is well described by the improved Bethe formula [50]. The residual energy loss in the software is used as an adjustable parameter to modify the improved Bethe formula, compensating for the average energy loss rate when the electron energy is lower than the work function of the sample [51]. CASINO treats the ionized valence electrons and the electrons excited by plasmons in the sample as secondary electrons. The fraction of ionized valence electrons in the sample can be calculated using Möller's equation [52].

Fig. 7 shows the three-dimensional model of the PET sample used in CASINO, with a length and width of 400 nm and a height of 100 nm. The electron beam energy was set to 2 keV, the incident angle to 4 °, and the angular divergence to 0.3 °. In CASINO, each incident electron is treated as an independent particle that does not affect others. Therefore, the absolute current of the beam is irrelevant in the simulation, and only the number of electrons needs to be set. For statistical purposes, we simulated 1 million electrons. Similarly, the beam diameter used in the simulation has no effect on the results. The beam diameter in the simulation was set to 1 nm. The calculation results are shown in Fig. 8. The deposited charges induced by incident electrons consists of the deposited incident electrons (Fig. 8(a),(b)) and the excited surface holes (Fig. 8(c),(d)). The deposition probability of incident electrons is 32.6%, and their positions are mainly within the depth range of 0 - 80 nm below the surface. The projection distribution of the intensity with respect to depth is shown in Fig. 8(b), with the intensity center at approximately 30 nm. The excitation probability of surface net holes is 77.19%, and their distribution depths are in the range of 0 - 1.5 nm, with greater intensity closer to the surface. The existence of a large number of holes on the surface results in a positively charged collision zone between the incident electrons and the surface. The deposition of incident electrons inside the PET makes the material negatively charged.

The attraction of surface positive charges to the incident electrons causes the transmission of low-energy electrons in the nanocapillaries to exhibit geometric transmission characteristics (Fig. 4), while electron scattering dominates[31] outside the geometric opening angle of the nanocapillaries. This prevents the electron transmission from having a guiding effect like that of positive ions. This is consistent with our current experimental measurements.

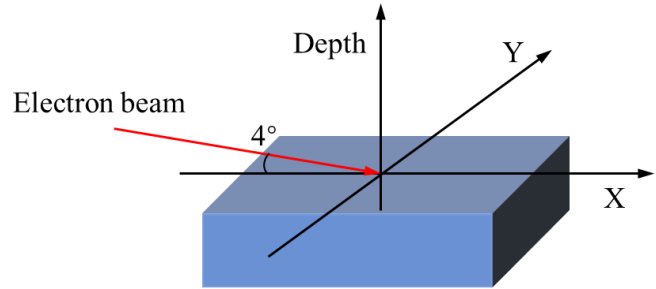

**Figure 7.** The 3-dimensional model of PET sample used in CASINO.

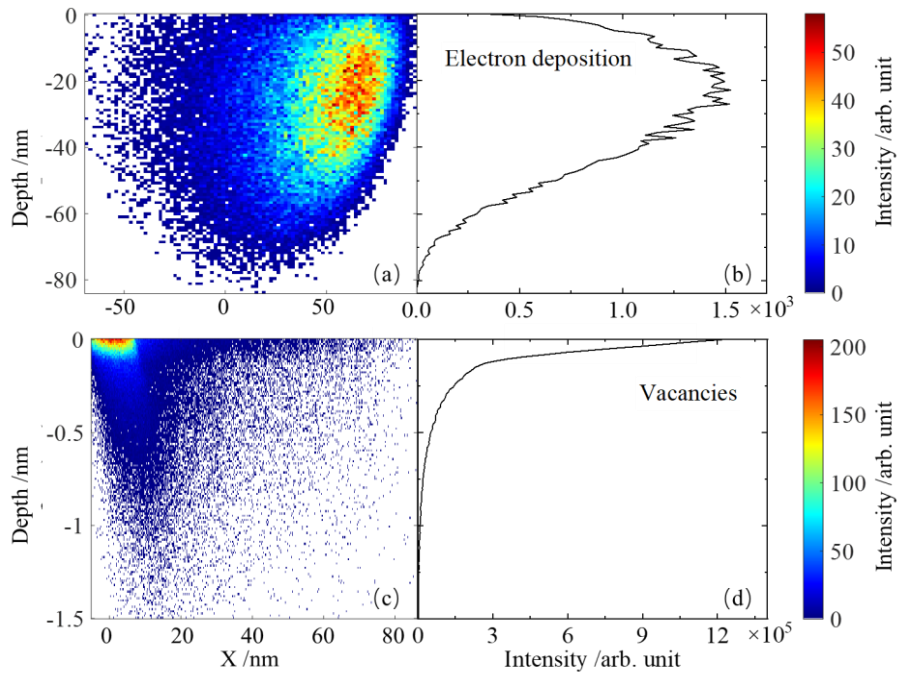

**Figure 8.** Calculation results of charge deposition on the surface of PET material caused by 2 keV electrons at an incidence angle of 4°: (a) The two-dimensional distribution of the deposited electrons and (b) its intensity distribution projection at depth; (c) the two-dimensional distribution of the holes, and (d) its intensity distribution projection at depth.

The center of the angular distribution of transmitted electrons on the plane of the observation angle $\phi$ moves opposite to that of the tilt angle of the nanocapillary (Fig. 3). This is due to the existence of positive charge spots on the inner wall where the electron beam directly collides, attracting the electron beam. When charging at a 0 ° tilt angle, the FWHM of the transmitted beam spot widens as charge accumulates (Fig. 5(b)), which is also caused by the attraction of positive charges on the inner wall. The oscillation phenomenon observed during the charging process (Fig.5 (a), Fig. 6) is due to the rapid recombination of surface positive charges and internal negative charges after they accumulate to a certain intensity, followed by their cyclical accumulation. Positive charges on the inner wall attract incident electrons to the surface, reducing their transmission probability and causing the transmitted electron spot to expand. Conversely, negative charges inside the body repel electrons entering the surface, increasing their exit probability and causing the transmitted electron spot to shrink. The competition between these effects results in complex charging oscillations.

For the phenomenon where the transmission angle distribution splits into two electron spots along the θ direction (Fig. 5(b), Fig. 6), it is considered that small-angle bending is inevitable when the membrane is fixed, as the thickness of the PET membrane is only 10 μm. As shown in Fig. 9, the PET membrane has a slight bend in the θ plane, which causes the electron beam to enter the nanocapillaries at a certain incident angle in the θ direction. For the upper part of the membrane, the incident electrons cause the upper surfaces of the nanocapillaries to become positively charged, attracting the electrons and causing them to deflect upward. For the lower part of the membrane, the incident electrons cause the lower surfaces of the nanocapillaries to become positively charged, attracting the electrons and causing them to deflect downward. As the charge accumulates, the positive charges on the upper and lower surfaces of the nanocapillaries gradually build up, increasing their attractive effect on the electrons. This causes the upper and lower electron spots to move farther apart until the charging reaches a stable state. After increasing the tilt angle, in the nanocapillaries on the upper part of the curved surface, the number of electrons colliding with the upper wall decreases, while the number of electrons colliding with the side wall increases. In the nanocapillaries on the lower part of the curved surface, fewer electrons collide with the lower wall, and more electrons collide with the side wall (the same side as the upper nanocapillaries). The charge distribution on the inner walls of the nanocapillaries in the upper and lower parts of the curved surface becomes similar, resulting in a single electron spot again.

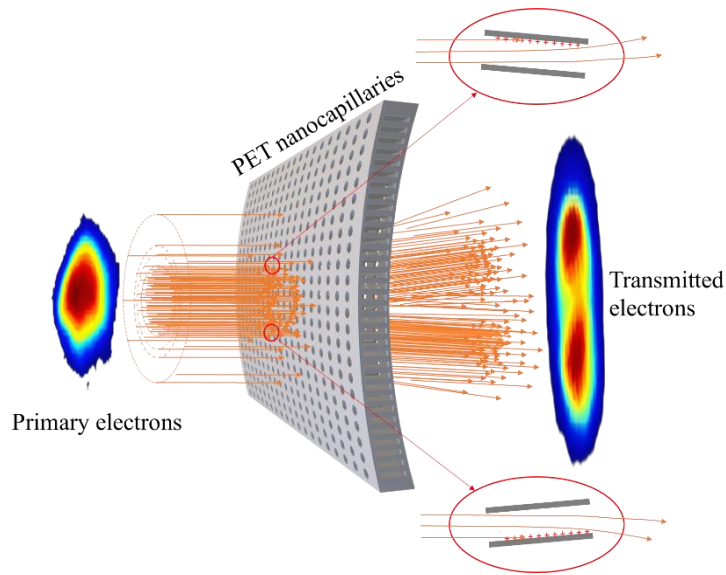

**Figure 9.** Schematic diagram of the distribution of transmitted electrons at 0° tilt angle.

The concave structure of the membrane can focus the ion beam, and the results of this paper show that the same concave structure may cause a splitting effect on the electron beam. The opposite transmission effects of the electron beam and ion beam will serve as powerful experimental evidence to prove the different charging mechanisms of the two, which is expected to resolve the controversy surrounding the charging mechanism of the electron beam. In the future, the author will precisely set the bending angle in the θ plane, conduct comparative experiments of the electron beam and ion beam, verify the hypothesis presented in this paper, and measure the differences in the electron and ion charging phenomena. At the same time, the results of this paper provide both data and theoretical support for the development of particle beam optics.

## 5. Conclusion

In this paper, the measurement results of 2 keV electron transport in 800 nm diameter PET nanocapillaries are presented. The results of the full angular distribution at steady state show that the center of the angular distribution of the transmitted electrons moves in the opposite direction to the tilt angle of the nanocapillaries, and no guiding effect is observed, as in the case of positive ions. The results of charging evolution at a 0° tilt angle show that the FWHM of the transmitted beam spot widens with the accumulation of charging time. An oscillation phenomenon exists during the charging process, and the peak position and FWHM of the transmitted electron spot oscillate synchronously with the transmission rate. CASINO was used to simulate the deposition of charges on the surface of PET. The simulation results showed that the surface of the nanocapillary was excited by electrons, producing a large number of net holes, thus forming a positive charge accumulation. Some of the incident electrons deposit deeper below the surface, forming a layer of negative charge. Based on this, the characteristic shift in the angular distribution center is explained as follows: there is a positive charge spot on the side where the electron beam directly collides with the inner surface of the nanocapillary, which attracts the electron beam. The incident electrons are drawn to the positive charges on the surface of the nanocapillary, resulting in the geometric transmission of the low-energy electrons. The electron scattering process dominates outside the geometric opening angle of the nanocapillary, which is not conducive to the generation of the guiding effect. The transmission broadening feature of the charging evolution is caused by the positive charge on the nanocapillary surface attracting electrons and expanding the transmitted electron spot. The oscillation is caused by the positive charge on the surface of the nanocapillary attracting the incident electrons, while the negative charge inside repels them, leading to repeated oscillations. The splitting of the transmitted electron angular distribution in the θ plane is considered to be caused by the bending of the membrane. A new experimental method is proposed to verify whether there is a difference between the charging mechanisms of electrons and ions. These results provide support for the development of electron transport control technology using insulating nanocapillaries.

# References

[1] Stolterfoht N, Bremer J H, Hoffmann V, Hellhammer R, Fink D, Petrov A, Sulik B 2002 *Phys. Rev. Lett.* **88** 133201

[2] Schiessl K, Palfinger W, Tökési K, Nowotny H, Lemell C, Burgdörfer J 2005 *Phys. Rev. A* **72** 062902

[3] Sahana M B, Skog P, Vikor G, Kumar R T R, Schuch R 2006 *Phys. Rev. A* **73** 040901（R

[4] Ikeda T, Kanai Y, Kojima T M, Iwai Y, Kambara T, Yamazaki Y, Hoshino M, Nebiki T, Narusawa T 2006 *Appl. Phys. Lett.* **89** 163502

[5] Skog P, Soroka I L, Johansson A, Schuch R 2007 *Nucl. Instr. Meth. Phys. Res. B* **258** 145

[6] Chen Y F, Chen X M, Lou F J, Xu J Z, Shao J X, Sun G Z, Wang J, Xi F Y, Yin Y Z, Wang X. A, Xu J K, Cui Y, Ding B W 2009 *Chin. Phys. B.* **18** 2739

[7] Cassimi A, Maunoury L, Muranaka T, Huber B, Dey K R, Lebius H, Lelièvre D, Ramillon J M, Been T, Ikeda T, Kanai Y, Kojima T M, Iwai Y, Yamazaki Y, Khemliche H, Bundaleski N, Roncin P 2009 *Nucl. Inst. Meth. Phys. Res. B* **267** 674

[8] Nakayama R, Tona M, Nakamura N, Watanabe H, Yoshiyasu N, Yamada C, Yamazaki A, Ohtani S, Sakurai M 2009 *Nucl. Inst. Meth. Phys. Res. B* **267** 2381

[9] Juhász Z, Sulik B, Rácz R, Biri S, Bereczky R J, Tőkési K, Kövér Á, Pálinkás J, Stolterfoht N 2010 *Phys. Rev. A* **82** 062903

[10] Skog P, Zhang H Q, Schuch R 2008 *Phys. Rev. Lett.* **101** 223202

[11] Zhang H Q, Skog P, Schuch R 2010 *Phys. Rev. A* **82** 052901

[12] Stolterfoht N, Hellhammer R, Sulik B, Juhász Z, Bayer V, Trautmann C, Bodewits E, Hoekstra R 2011 *Phys. Rev. A* **83** 062901

[13] Zhang H Q, Akram N, Skog P, Soroka I L, Trautmann C, Schuch R 2012 *Phys. Rev. Lett.* **108** 193202

[14] Zhang H Q, Akram N, Soroka I L, Trautmann C, Schuch R 2012 *Phys. Rev. A* **86** 022901

[15] Zhang H Q, Akram N, Schuch R 2016 *Phys. Rev. A* **94** 032704

[16] Giglio E, Guillous S, Cassimi A, Zhang H Q, Nagy G U L, Tőkési K 2017 *Phys. Rev. A* **95** 030702（R

[17] Giglio E, Guillous S, Cassimi A 2018 *Phys. Rev. A* **98** 052704

[18] Lemell C, Burgdörfer J, Aumayr F 2013 *Prog. Surf. Sci.* **88** 237

[19] Stolterfoht N, Yamazaki Y 2016 *Phys. Rep.* **629** 1

[20] Wan C L, Pan Y Z, Zhu L P, Zhang H W, Zhao Z Y, Yuan H, Li P F, Fan X H, Sun W S, Du Z H, Chen Q, Cui Y, Liao T F, Wei X H, Wang T Q, Chen X M, Li G P, Schuch R, Zhang H Q 2024 *Acta Phys. Sin.* **73** 104101

[21] Sun G Z, Chen X M, Wang J, Chen Y F, Xu J K, Zhou C L, Shao J X, Cui Y, Ding B W, Yin Y Z, Wang X A, Lou F J, Lv X Y, Qiu X Y, Jia J J, Chen L, Xi F Y, Chen Z C, Li L T, Liu Z Y 2009 *Phys. Rev. A* **79** 052902

[22] Chen L, Guo Y L, Jia J J, Zhang H Q, Cui Y, Shao J X, Yin Y Z, Qiu X Y, Lv X Y, Sun G Z, Wang J, Chen Y F, Xi F Y, Chen X M 2011 *Phys. Rev. A* **84** 032901

[23] Chen L, Lv X Y, Jia J J, Ji M C, Zhou P, Sun G Z, Wang J, Chen Y F, Xi F Y, Cui Y, Shao J X, Qiu X Y, Guo Y L, Chen X M 2011 *Phys. B: At. Mol. Opt. Phys.* **44** 045203

[24] Zhang Q, Liu Z L, Li P F, Jin B, Song G Y, Jin D K, Niu B, Wei L, Ha S, Xie Y M, Ma Y, Wan C L, Cui Y, Zhou P, Zhang H Q, Chen X M 2018 *Phys. Rev. A* **97** 042704

[25] Ha S, Zhang W M, Xie Y M, Li P F, Jin B, Niu B, Wei L, Zhang Q, Liu Z L, Ma Y, Lu D, Wan C L, Cui Y, Zhou P, Zhang H Q, Chen X M 2020 *Acta Phys. Sin.* **69** 094101

[26] Liu Z L, Ha S, Zhang W M, Xie Y M, Li P F, Jin B, Zhang Q, Ma Y, Lu D, Wan C L, Cui Y, Zhou P, Zhang H Q, Chen X M 2021 *Nucl. Phys. Rev.* **38** 95

[27] Milosavljević A R, Víkor Gy, Pešić Z D, Kolarž P, Šević D, Marinković B P, Mátéfi-Tempfli S, Mátéfi-Tempfli M, Piraux L 2007 *Phys. Rev. A* **75** 030901（R

[28] Milosavljević A R, Jureta J, Víkor G, Pešić Z D, Šević D, Mátéfi-Tempfli M, Mátéfi-Tempfli S, Marinković B P 2009 *Europhys. Lett.* **86** 23001

[29] Milosavljević A, Schiessl K, Lemell C, Mátéfi-Tempfli M, Mátéfi-Tempfli S, Marinković B P, Burgdörfer J 2012 *Nucl. Instr. Meth. Phys. Res. B* **279** 190.

[30] Das S, Dassanayake B S, Winkworth M, Baran J L, Stolterfoht N, Tanis J A 2007 *Phys. Rev. A* **76** 042716

[31] Schiessl K, Tőkési K, Solleder B, Lemell C, Burgdörfer J 2009 *Phys. Rev. Lett.* **102** 163201

[32] Dassanayake B S, Keerthisinghe D, Wickramarachchi S, Ayyad A, Das S, Stolterfoht N, Tanis J A 2013 *Nucl. Instr. Meth. Phys. Res. B* **298** 1

[33] Keerthisinghe D, Dassanayake B S, Wickramarachchi S J, Stolterfoht N, Tanis J A 2013 *Nucl. Instr. Meth. Phys. Res. B* **317** 105

[34] Keerthisinghe D, Dassanayake B S, Wickramarachchi S J, Stolterfoht N, Tanis J A 2016 *Nucl. Instr. Meth. Phys. Res. B* **382** 67

[35] Keerthisinghe D, Dassanayake B S, Wickramarachchi S J, Stolterfoht N, Tanis J A 2015 *Phys. Rev. A* **92** 012703

[36] Vokhmyanina K A, Kubankin A S, Myshelovka L V, Zhang H Q, Kaplii A A, Sotnikova V S, Zhukova M A 2020 *J. Instrum.* **15** C04003

[37] Dassanayake B S, Das S, Bereczky R J, Tőkési K, Tanis J A 2010 *Phys. Rev. A* **81** 020701（R

[38] Dassanayake B S, Bereczky R J, Das S, Ayyad A, Tökési K, Tanis J A 2011 *Phys. Rev. A* **83** 012707

[39] Wickramarachchi S J, Ikeda T, Dassanayake B S, Keerthisinghe D, Tanis J A 2016 *Phys. Rev. A* **94** 022701

[40] Stolterfoht N, Tanis J A 2018 *Nucl. Instr. Meth. Phys. Res. B* **421** 32

[41] Wan C L, Li P F, Qian L B, Jin B, Song G Y, Gao Z M, Zhou L H, Zhang Q, Song Z Y, Yang Z H, Shao J X, Cui Y, Schuch R, Zhang H Q, Chen X M 2016 *Acta Phys. Sin.* **65** 204103

[42] Qian L B, Li P F, Jin B, Jin D K, Song G Y, Zhang Q, Wei L, Niu B, Wan C L, Zhou C L, Müller A M, Dobeli M, Song Z Y, Yang Z H, Schuch R, Zhang H Q, Chen X M 2017 *Acta Phys. Sin.* **66** 124101

[43] Li P F, Yuan H, Cheng Z D, Qian L B, Liu Z L, Jin B, Ha S, Wan C L, Cui Y, Ma Y, Yang Z H, Lu D, Schuch R, Li M, Zhang H Q, Chen X M 2022 *Acta Phys. Sin.* **71** 074101

[44] Li P F, Yuan H, Cheng Z D, Qian L B, Liu Z L, Jin B, Ha S, Zhang H W, Wan C L, Cui Y, Ma Y, Yang Z H, Lu D, Schuch R, Li M, Zhang H Q, Chen X M 2022 *Acta Phys. Sin.* **71** 084104

[45] Zhou P, Wan C L, Yuan H, Cheng Z D, Li P F, Zhang H W, Cui Y, Zhang H Q, Chen X M 2023 *High Pow. Laser Part. Beams* **35** 026001

[46] Data sheets of Mylar（http://www.dupontteijinfilms.com

[47] Hovington P, Drouin D, Gauvin R. 1997 *Scanning* **19** 1

[48] Drouin D, Couture A R, Joly D, Taster X, Aimez V, Gauvin R 2007 *Scanning* **29** 92

[49] Demers H, Poirrier-Demers N, Couture A R, Joly D, Guilmain M, Jonge N D, Drouin D 2011 *Scanning* **33** 135

[50] Joy D C, Luo S 1989 *Scanning* **11** 176

[51] Lowney J R 1996 *Scanning* **18** 301

[52] Reimer L 1998 *Scanning Electron Microscopy*（2nd Ed.）（Berlin: Springer）pp57–134